\documentclass[10pt,letterpaper,twocolumn]{article} 
\usepackage{ol2}
\usepackage{amsmath}
\begin{document}
\twocolumn[ 

\title{Add-drop system for Kerr bistable memory in silicon nitride microrings}
\author{Wataru Yoshiki and Takasumi Tanabe$^{*}$}
\address{
Department of Electronics and Electrical Engineering, Faculty of Science and Technology, Keio University, \\ 3-14-1, Hiyoshi, Kohoku-ku, Yokohama 223-8522, Japan\\
$^*$Corresponding author: takasumi@elec.keio.ac.jp
}

\begin{abstract}
 We numerically studied the required conditions for achieving a memory operation based on Kerr bistability in the presence of the thermo-optic effect in silicon nitride microring cavity systems.  By comparing a side-coupled system and an add-drop system, we found that the latter is needed to realize obtain Kerr bistable operation.  In addition, we show the coupling coefficient range in which the memory operation does not suffer from the thermo-optic effect.
\end{abstract}
 ] 
 
 \noindent
 Optical bistability is an important physical phenomenon for achieving optical memories, optical transistors, and logic gates, which are the building blocks of all-optical signal processing.  Recent progress on the technology used to fabricate optical microcavities\cite{Armani2003uhq,Pollinger2010aos,Kudo2013fwg} has enabled us to demonstrate optical bistable behavior at a very low input power\cite{Nozaki2012upa,Pollinger2010aos}. 
 The simplest way to demonstrate optical bistability in a microcavity is to modulate the refractive index of the cavity, and generally three effects such as a thermo-optic (TO) effect\cite{Almeida2004obs}, a carrier plasma effect\cite{Tanabe2005fba,Kawashima2008obr} and a Kerr effect\cite{Pollinger2010aos,Razdolskiy2011hmn,Eckhouse2012kia}, are employed for this purpose. 
 Of these, the Kerr effect is regarded as the most promising due to its instantaneous response and low loss.
 To take advantage of its excellent features, researchers have already investigated optical switches and memories based on Kerr effect both theoretically\cite{Soljacic2002obs,Yanik2003hca,Shafiei2010ltb,Yoshiki2012abm} and experimentally\cite{Pollinger2010aos,Razdolskiy2011hmn,Eckhouse2012kia}.
 
 In recent years, the silicon nitride ($\mathrm{Si_3N_4}$) microring cavity\cite{Gondarenko2009hcm} has attracted a lot of attention as a prominent platform for nonlinear optical processes\cite{Levy2010ccm,Okawachi2011osf} because it has a high quality factor and a large band gap which suppress multi-photon absorption.
 It is also an excellent on which to demonstrate a Kerr bistable memory\cite{Ikeda2008tak}. 
 However, the refractive index change induced by the Kerr effect is generally very small compared with that induced by the TO and the carrier-plasma effects\cite{Uesugi2006ion,Barclay2005nrs}, and it appears to be difficult to use the Kerr effect without suffering from other effects. 
 It is shown experimentally that $\mathrm{Si_3N_4}$ has relatively large absorption ($0.055~\mathrm{dB/cm}$\cite{Gondarenko2009hcm} compared with $0.2~\mathrm{dB/km}$\cite{Miya1979ull} of silica), so it remains an open question whether or not a  Kerr bistable memory is feasible in a $\mathrm{Si_3N_4}$ microring cavity when the TO effect is present. (Note that the carrier-plasma effect is negligible in $\mathrm{Si_3N_4}$ because of the large bandgap).
 
 We are also interested in whether or not a simple side-coupled configuration is sufficient to obtain bistable memory operation in the presence of TO effect.  It has been shown theoretically that a side-coupled system exhibits high-contrast bistability when the cavity is made of an ideal Kerr medium\cite{Yanik2003hca}.  But we feel that further analysis is needed if we are to find a good design for the experimental demonstration.  Indeed we will show later in this letter that it is not possible to demonstrate optical bistable memory with a side-coupled cavity in a realistic material.
 
 With these points in mind, we studied the required conditions numerically and determined the cavity configuration and the range of optimal couplings needed to obtain Kerr bistable memory in $\mathrm{Si_3N_4}$ microring cavities in the presence of the TO effect.
 
\begin{figure}[htbp]
  \begin{center}
   \includegraphics*[width=3.2in]{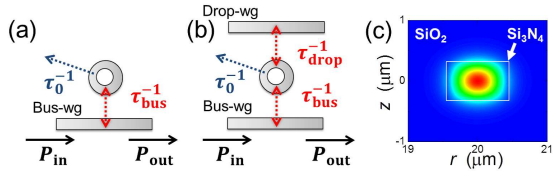}
   \caption{(color online) Schematic image of (a) a side-coupled system and (b) an add-drop system. (c) Cross-sectional mode profile of the microring cavity.}
   \label{fig:CMT1}
  \end{center}
 \end{figure}

 Figures~\ref{fig:CMT1}(a) and (b) are the models we used for the analysis.  There are two possible ring resonator system configurations: One is a side-coupled system (Fig.~\ref{fig:CMT1}(a)) and the other is an add-drop system  (Fig.~\ref{fig:CMT1}(b)).
 The master equations of our analysis are derived from coupled mode theory equations\cite{Yoshiki2012abm, Manolatou1999cma} by substituting $d/dt=0$ (steady state) and are given as,
 \begin{eqnarray}
  &P_\mathrm{in}& = \tau_\mathrm{bus} \left[ \Delta \omega^2 + \frac{1}{4\tau_\mathrm{load}^2} \right]U_\mathrm{cavity} \label{eq:master1} \\
  &P_\mathrm{out}& = \tau_\mathrm{bus} \left[ \Delta \omega^2 \right. \nonumber \\
  && \left. + \left( \frac{1}{2\tau_\mathrm{load}} - \frac{1}{\tau_\mathrm{bus}} \right)^2 \right] U_\mathrm{cavity} \label{eq:master2} \\
  &\Delta \omega& = 2\pi c \left[ \frac{1}{\lambda_0 \left(1+\frac{\Delta n_\mathrm{Kerr} + \Delta n_\mathrm{TO}}{n_\mathrm{eff}}\right)} - \frac{1}{\lambda} \right] \\
  &\Delta n_\mathrm{Kerr}& = \frac{c \int\!\!\!\int n_2(r,z) U_\mathrm{cavity}  \tilde I^2(r,z) \mathrm{d}r \mathrm{d}z}{\pi n_\mathrm{eff} R \int\!\!\!\int \tilde I(r,z) \mathrm{d}r \mathrm{d}z}, \label{eq:master3}
 \end{eqnarray}
 where $P_\mathrm{in}$, $P_\mathrm{out}$, $U_\mathrm{cavity}$, $\Delta n_\mathrm{Kerr}$ and $\Delta n_\mathrm{TO}$ are the input power, the output power, the light energy stored in the cavity and the refractive index changes induced by the Kerr and TO effects, respectively.
 $c$, $n_\mathrm{eff}$, $n_2(r,z)$,  $\lambda_0$, $\lambda$, $R$ and $\tilde I(r,z)$  are the velocity of light, the effective refractive index of the cavity, the nonlinear refractive index of the cavity, the resonant wavelength of the cavity, the input wavelength, the cavity radius and the normalized intensity profile (Fig.~\ref{fig:CMT1}(c)), respectively.  
 $\tau_\mathrm{load}$ is the loaded photon lifetime, given as $\tau_\mathrm{load}^{-1} = \tau_0^{-1} + \tau_\mathrm{bus}^{-1} + \tau_\mathrm{drop}^{-1}$, where $\tau_0$, $\tau_\mathrm{bus}$ and $\tau_\mathrm{drop}$ are the intrinsic photon lifetime, and the coupling photon lifetime between the cavity and the bus/drop waveguides, respectively (Fig.~\ref{fig:CMT1}~(b)).
 Note that we can use the same equations for a side-coupled system by setting $\tau_\mathrm{drop}$ at infinity.
 
 We employ a $\mathrm{Si_3N_4}$ microring cavity with silica cladding and whose radius, thickness and waveguide width are $20~\mathrm{\mu m}$, $644~\mathrm{nm}$ and $900~\mathrm{nm}$, respectively.
 Its intrinsic photon lifetime $\tau_\mathrm{0}$ is $2.46~\mathrm{ns}$, which is derived from the photon lifetime determined by the absorption loss $\tau_\mathrm{abs} = 4.4~\mathrm{ns}$ and the photon lifetime determined by the scattering loss $\tau_\mathrm{scat} = 5.58~\mathrm{ns}$.
 These parameters are taken from Ref.~\onlinecite{Gondarenko2009hcm}.
 
 It is well known that wavelength detuning $\delta \lambda= \lambda - \lambda_0$ determines the eye-openings of a hysteresis curve and needs to be set larger than $(\sqrt{3}/2)\lambda_\mathrm{FWHM}$ \cite{Yanik2003hca}, where $\lambda_\mathrm{FWHM}$ is the full-width half-maximum of the cavity resonance.  
 In addition, $\tau_\mathrm{bus}$ and $\tau_\mathrm{drop}$ are also important for memory operation.  As described above, the Kerr effect is much smaller but faster than the TO effect; hence we have to finish the operation before the TO effect becomes apparent when we want to demonstrate a Kerr bistable memory.  For this, we have to adopt a small $\tau_\mathrm{load} = (\tau_\mathrm{0}^{-1} + \tau_\mathrm{bus}^{-1} + \tau_\mathrm{drop}^{-1})^{-1}$ to increase the response speed of the cavity.  $\tau_\mathrm{bus}$ and $\tau_\mathrm{drop}$ are the design parameters.  These parameter can be changed by varying the distance between the cavity and waveguides however $\tau_0$ is fixed by the material and structure of the cavity.  Therefore, adopting small value for $\tau_\mathrm{bus}$ and $\tau_\mathrm{drop}$ is the simplest way to finish the operation quickly.  However we will show later that the strategy is not that simple.
 
 So the problem is to find the acceptable ranges for $\tau_\mathrm{bus}$, $\tau_\mathrm{drop}$, and $\delta$ that satisfy the following two conditions:
 
 Condition~(1): The on-off contrast between two bistable output states is sufficiently large.
 
 Condition~(2): The TO effect is minimized until the end of the memory operation.
 
\begin{figure}[htbp]
  \begin{center}
   \includegraphics*[width=2.4in]{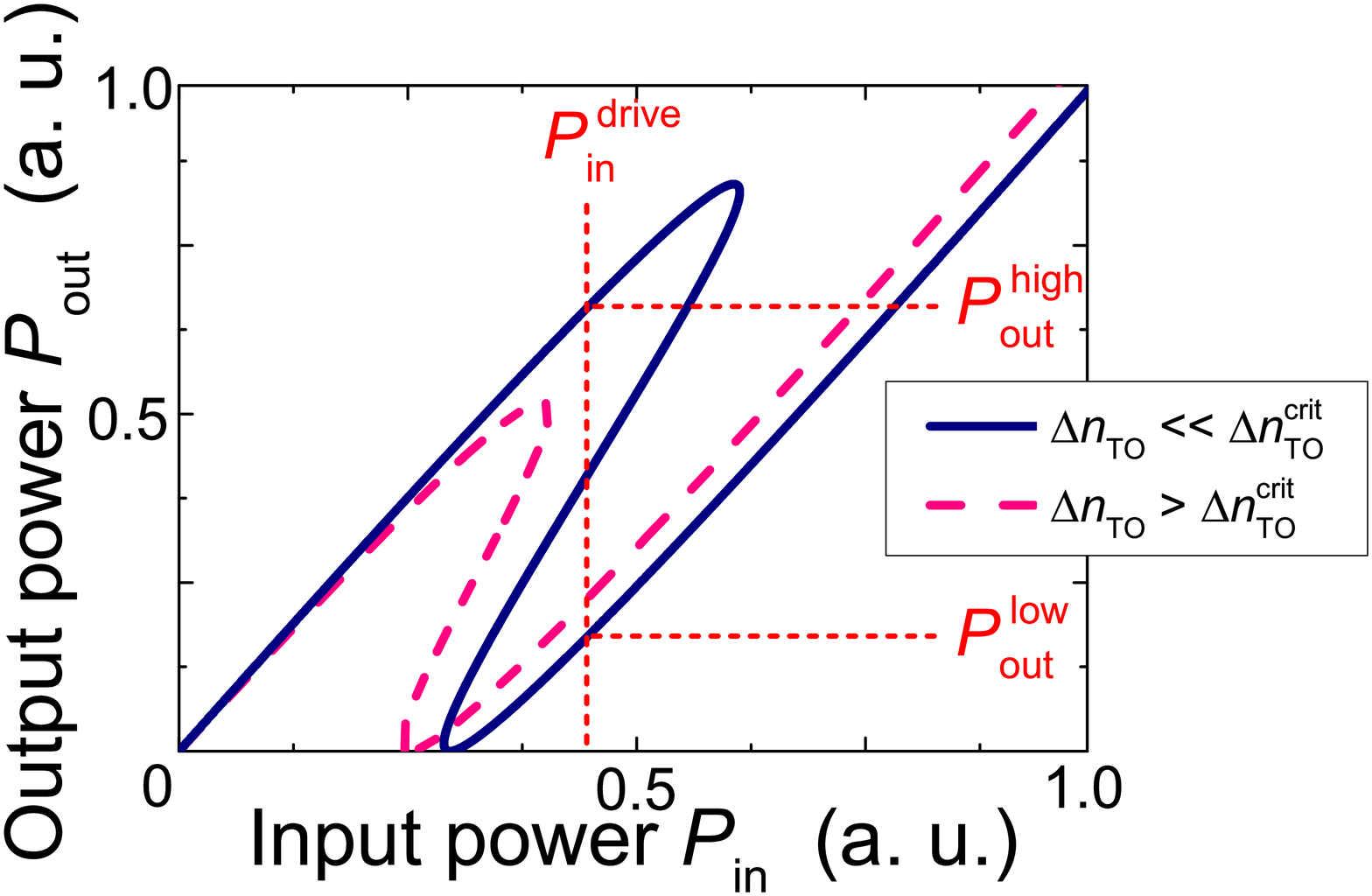}
   \caption{(color online) Input $P_\mathrm{in}$ versus output $P_\mathrm{out}$ for a side-coupled cavity at $\delta \lambda =(\sqrt{3}/2) \lambda_\mathrm{FWHM}$ and $\tau_\mathrm{bus} = \tau_\mathrm{0}$ (critical coupling condition).  The solid line indicates no TO effect is present.  The dashed line shows $\Delta n_\mathrm{TO} > \Delta n_\mathrm{TO}^\mathrm{crit}$.}
   \label{fig:bistability1}
  \end{center}
 \end{figure}

 First, condition (1) is studied by using Eqs.~(\ref{eq:master1})--(\ref{eq:master3}).
 When we consider an ideal Kerr medium and neglect other nonlinearities, the input ($P_\mathrm{in}$) and output ($P_\mathrm{out}$) powers for a side-coupled system exhibit hysteresis behavior as shown by the solid line in Fig.~\ref{fig:bistability1}. 
 To achieve bistable memory operation, the input drive power is set at $P_\mathrm{in}^\mathrm{drive} = \left(P_\mathrm{in}^\mathrm{set} + P_\mathrm{in}^\mathrm{reset}\right)/2$, where $P_\mathrm{in}^\mathrm{set}$ and $P_\mathrm{in}^\mathrm{reset}$ are the upper and lower bounds of the input power of the hysteresis eye-opening.
 The output power on the input drive power $P_\mathrm{in}^\mathrm{drive}$ is given as $P_\mathrm{out}^\mathrm{high}$ and $P_\mathrm{out}^\mathrm{low}$ as shown in Fig.~\ref{fig:bistability1}.
 We define $\eta = \left(P_\mathrm{out}^\mathrm{high} - P_\mathrm{out}^\mathrm{low}\right)/P_\mathrm{out}^\mathrm{high}$ as the normalized contrast between two bistable output states.
 
\begin{figure}[htbp]
  \begin{center}
   \includegraphics*[width=3.2in]{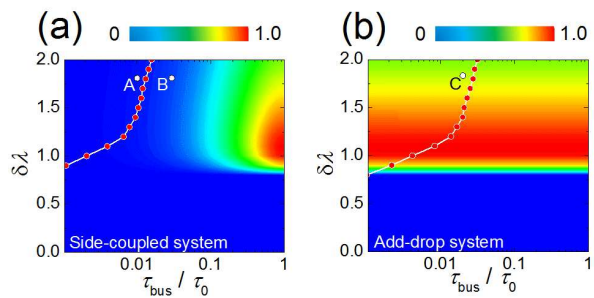}
   \caption{(color online) $\eta$ for different $\tau_\mathrm{bus}$ and $\delta \lambda$ values.  The red dots are when $\Delta n_\mathrm{TO} = \Delta n_\mathrm{TO}^\mathrm{crit}$. (a) A side-coupled system. (b) An add-drop system.}
   \label{fig:colormap1}
  \end{center}
 \end{figure}

 We calculate $\eta$ for different $\delta \lambda$ and $\tau_\mathrm{bus}$ values.  The results are shown as an intensity map in Fig.~\ref{fig:colormap1}(a) for a side coupled cavity and Fig.~\ref{fig:colormap1}(b) for an add-drop system.  In an add-drop system (Fig.~\ref{fig:colormap1}(b)) $\tau_\mathrm{drop}$ is properly controlled to satisfy the critical coupling condition $(\tau_\mathrm{drop}^{-1} = \tau_\mathrm{bus}^{-1} - \tau_\mathrm{0}^{-1})$.   Figure~\ref{fig:colormap1}(a) indicates that the contrast disappears in a side-coupled system when $\tau_\mathrm{bus}$ is small.  This is because the cavity and the waveguide are strongly over-coupled and the dip at the resonance is shallow.  It should be noted that we cannot demonstrate a bistable memory with this condition because we cannot distinguish on and off.
 On the other hand, Fig.~\ref{fig:colormap1}(b) shows that $\eta$ is large even when $\tau_\mathrm{bus}$ is small in an add-drop system because the critical coupling condition is satisfied by appropriately controlling $\tau_\mathrm{drop}$.
 
 Next, condition (2) is studied.  When the cavity is charged with photons, heat accumulates as a result of light absorption.  After a certain period, $\Delta n_\mathrm{TO}$ becomes larger than a critical value $\Delta n_\mathrm{TO}^\mathrm{crit}$, and the Kerr memory operation will fail.  The dashed line in Fig.~\ref{fig:bistability1} shows the $P_\mathrm{out}$ versus $P_\mathrm{in}$ when $\Delta n_\mathrm{TO} > \Delta n_\mathrm{TO}^\mathrm{crit}$.  Unlike when no TO effect is present (solid line), the hysteresis curve deforms.  There is only one output state for the input drive power $P_\mathrm{in}^\mathrm{drive}$.  $P_\mathrm{in}^\mathrm{drive}$ is the constant drive power input that we add during the entire memory operation.  Namely $\Delta n_\mathrm{TO}^\mathrm{crit}$ is the refractive index change when the output states for $P_\mathrm{in}^\mathrm{drive}$ becomes one and the optical bistable memory operation fails.
 
 We next calculate $\Delta n_\mathrm{TO}^\mathrm{end}$, namely the $\Delta n_\mathrm{TO}$ value at the end of the memory operation, by using transient analysis, which is similar to the approach described in Ref.~\onlinecite{Yoshiki2012abm}.
 Note that $\Delta n_\mathrm{TO}$ is calculated with a thermal diffusion equation that assumes a heat source distribution given by Fig.~\ref{fig:CMT1}(c).
 The duration of the memory operation is set at $70 \tau_\mathrm{load}$, which is sufficiently long for the system to approach equilibrium.
 The memory is turned on and off by employing set and reset pulses with powers of $2P^\mathrm{drive}_\mathrm{in}$ and 0 (negative pulse), with a duration of $5\tau_\mathrm{load}$ at timings of $t=20\tau_\mathrm{load}$ and $t=40\tau_\mathrm{load}$.
 Now, by comparing $\Delta n_\mathrm{TO}^\mathrm{crit}$ and $\Delta n_\mathrm{TO}^\mathrm{end}$ for different $\delta \lambda$ and $\tau_\mathrm{bus}$ values, we know whether or not the Kerr memory operation will work during the entire memory operation time.
 
 The red dots in Fig.~\ref{fig:colormap1} indicate the conditions when $\Delta n_\mathrm{TO}^\mathrm{end} = \Delta n_\mathrm{TO}^\mathrm{crit}$.
 In the area to the left of the plots is $\Delta n_\mathrm{TO}^\mathrm{end} < \Delta n_\mathrm{TO}^\mathrm{crit}$, where the Kerr bistable memory works during the entire operation time without suffering from the TO effect.  In contrast, the area to the right of the plots is the $\Delta n_\mathrm{TO}^\mathrm{end} > \Delta n_\mathrm{TO}^\mathrm{crit}$ regime, where the TO effect becomes dominant before the completion of the memory operation.  Therefore, we can only use the left area for Kerr bistable memory operation.  This is condition (2).
 
 Now, conditions (1) and (2) are considered simultaneously.  Although $\eta$ is large for Fig.~\ref{fig:colormap1}(b) even in a $\Delta n_\mathrm{TO}^\mathrm{end} < \Delta n_\mathrm{TO}^\mathrm{crit}$ regime, it is nearly zero for a side-coupled system (Fig.~\ref{fig:colormap1}(a)).
 This result indicates that it is difficult to obtain a Kerr bistable memory in the presence of a TO effect if we employ a side-coupled system.  We would like to emphasize that even though it has been predicted that a Kerr bistable memory will obtain high-contrast switching in an ideal Kerr medium for a side-coupled cavity system \cite{Yanik2003hca}, this is not true for a material where a TO effect is present.  Indeed, our analysis revealed that we need to employ an add-drop system.  This is because we have additional controllability with the drop waveguide that enables the cavity to exhibit critical coupling even when the bus waveguide is strongly over-coupled with the cavity.
 
 \begin{figure}[htbp]
  \begin{center}
   \includegraphics*[width=3.2in]{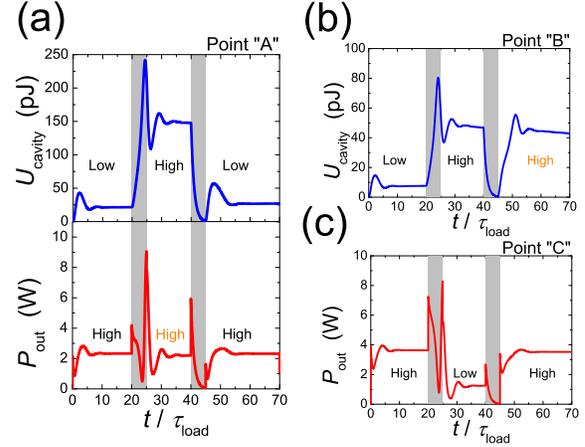}
   \caption{(color online) Time domain response of memory at points ``A'', ``B'' and ``C'' in Fig.~\ref{fig:colormap1}. (a) The light energy $U_\mathrm{cavity}$ and the output power $P_\mathrm{out}$ at point ``A'', (b) $U_\mathrm{cavity}$ at point ``B'' and (c) $P_\mathrm{out}$ at point ``C'' are shown. Set and reset pulses are inputted in the gray zone. The input drive powers $P_\mathrm{in}^\mathrm{drive}$ are $2.2~\mathrm{W}$, $257~\mathrm{mW}$ and $3.9~\mathrm{W}$.}
   \label{fig:colormap2}
  \end{center}
 \end{figure}
 
 To verify the understanding obtained from Fig.~\ref{fig:colormap1}(a) and (b), we calculate the time domain responses of the memory.
 Figure~\ref{fig:colormap2}(a)--(c) show the time domain Kerr memory response at points ``A'' ($\tau_\mathrm{bus}/\tau_\mathrm{0}=0.01$ and $\delta \lambda = 1.8$), ``B'' ($\tau_\mathrm{bus}/\tau_\mathrm{0}=0.03$ and $\delta \lambda = 1.8$) and ``C'' ($\tau_\mathrm{bus}/\tau_\mathrm{0}=0.02$ and $\delta \lambda = 1.8$) in Figs.~\ref{fig:colormap1}(a) and (b).
 At point ``A'', $U_\mathrm{cavity}$ shows a clear bistable memory operation but $P_\mathrm{out}$ does not (Fig.~\ref{fig:colormap2}(a)).
 Since ``A'' is in the $\Delta n_\mathrm{TO}^\mathrm{end} < \Delta n_\mathrm{TO}^\mathrm{crit}$ regime, the TO effect is minimum.  Hence  $U_\mathrm{cavity}$ exhibit bistable behavior.
 However, due to the failure of condition (1) (i.e. the system is strongly over-coupled and $\eta$ is nearly zero), we cannot distinguish two bistable output states.
 Point ``B'' is in the $\Delta n_\mathrm{TO}^\mathrm{end} > \Delta n_\mathrm{TO}^\mathrm{crit}$ regime, hence we cannot observe any bistable behavior for $P_\mathrm{out}$ or even for $U_\mathrm{cavity}$ (Fig.~\ref{fig:colormap2}(b)).
 This is because $\Delta n_\mathrm{TO}$ reaches $\Delta n_\mathrm{TO}^\mathrm{crit}$ at $t = 35 \tau_\mathrm{load}$ and $U_\mathrm{cav}$ cannot maintain bistability.
 Unlike the side coupled system, we successfully obtain bistable behavior in $P_\mathrm{out}$ (Fig.~\ref{fig:colormap2}(c)) at point ``C'', because conditions (1) and (2) are both satisfied ($\Delta n_\mathrm{TO}^\mathrm{end} < \Delta n_\mathrm{TO}^\mathrm{crit}$ and large $\eta$).
 From these results, we conclude that the understandings obtained from Fig.~\ref{fig:colormap1}(a) and (b) are truly valid.
 
 Finally, we briefly discuss the driving power of the memory.
 As seen in Fig.~\ref{fig:colormap2}(c), we need to input a few watts to drive the memory.
 This value is very high; in fact, it is higher than recently reported memories demonstrated with different types of cavities \cite{Yoshiki2012abm,Pollinger2010aos}.
 In accordance with Ref.~\onlinecite{Notomi2011lpn}, the required input power $P_\mathrm{in}^\mathrm{req}$ is described as,
 \begin{eqnarray}
  P_\mathrm{in}^\mathrm{req}  = \frac{\varepsilon n_\mathrm{0} \omega}{2n_2}\frac{V_\mathrm{cav}}{Q_\mathrm{load}^2},
   \label{eq:inputpower1}
 \end{eqnarray}
 where $n_0$, $\varepsilon$, $\omega$, $V_\mathrm{cav}$ and $Q_\mathrm{load}$ are the refractive index of the medium, the permittivity of the medium, the angular frequency, the cavity mode volume and the loaded $Q$, respectively.
 Equation~(\ref{eq:inputpower1}) indicates that $P_\mathrm{in}^\mathrm{req}$ is inversely proportional to the square of $Q_\mathrm{load}$, so a simple way to reduce the input power is to use a cavity with higher $Q$ than $\mathrm{Si_3N_4}$ microring.
 Silica toroid microcavities\cite{Armani2003uhq,Yoshiki2012abm} and silica bottle cavities\cite{Pollinger2010aos} are good candidates because they have ultra high $Q$ values $10^8$.  Indeed, we performed the same analysis by using a silica toroid microcavity and obtained the same results but with much smaller driving power of less than $10~\mathrm{mW}$ \cite{Yoshiki2012abm}.
 
 In conclusion, we provided a detailed analysis of a Kerr bistable memory in a $\mathrm{Si_3N_4}$ microring cavity and provided the conditions required for achieving a Kerr bistable memory in the presence of a TO effect.
 As a result of the analysis, we concluded that a strongly over-coupled add-drop configuration is needed to achieve a Kerr bistable memory in a $\mathrm{Si_3N_4}$ microring, of which conclusion is different from that obtained with Ref.~\onlinecite{Yanik2003hca} where authors assumed an ideal Kerr medium.
 Our study reveals the potential of the $\mathrm{Si_3N_4}$ microring as a platform for a Kerr bistable memory, fills up the gap between theory and experiment, and provides a clear strategy for upcoming experiments.
 
 This work is supported in part by the Strategic Information and Communications R\&D Promotion Programme (SCOPE) of the Ministry of Internal Affairs and Communications in Japan, the Toray Science Foundation, the Support Center for Advanced Telecommunications Technology Research, Foundation, and the Leading Graduate School program for ``Science for Development of Super Mature Society'' from the Ministry of Education, Culture, Sports, Science, and Technology in Japan.

\bibliographystyle{osajnl}{

\end{document}